# Pre-fetching tree-structured data in distributed memory

Lex Weaver and Chris Johnson
E-mail: Chris.Johnson@cs.anu.edu.au
Department of Computer Science
Australian National University
Canberra, A.C.T. 0200 Australia
phone: +61 6 249 2624
fax: +61 6 249 0010

## Abstract

A distributed heap storage manager has been implemented on the Fujitsu AP1000 multicomputer. The performance of various pre-fetching strategies is experimentally compared. Subjective programming benefits and objective performance benefits of up to 10% in pre-fetching are found for certain applications, but not for all. The performance benefits of pre-fetching depend on the specific data structure and access patterns. We suggest that control of pre-fetching strategy be dynamically under the control of the application.

## 1 Introduction

In many applications in parallel distributed computing a lot of programmer effort is expended in managing distributed and shared data structures. In many cases the structures are essential to the correct functioning of the program, but their performance is not critical to the overall program performance. Programming the data sharing is often a distraction from the primary skills of the programmer, and the central algorithm, which may be in visualisation and rendering [1], or scientific calculations such as the N-body problem [3, 4]. Providing a generic *shared heap* library therefore has the potential to ease the programmer's job, providing the performance impact is acceptable.

We have implemented a shared heap utility on the Fujitsu AP1000 multicomputer. The *shared heap* nominates one or more processor cells as servers, and allows all other cells to do general computations and to act as clients of the server cells.

The client-side has a library of client routines for creating, reading and updating structured data that we refer to as *objects*. The server-side has one process (task) on each server cell which acts as the object manager for a certain group of objects. As more cells are designated as managers a larger total space becomes available for use as object storage.

The design issues in creating such a system include:

- whether objects can be referenced by more than one name

- whether data is cached in client memory

- if data is cached, how space is recovered from a full cache

- data blocking: that is, how data is passed around, as whole objects, part objects, or storage units predicted single objects, sets of objects, or storage units likewise

- whether data is fetched on demand or may be predictively pre-fetched

- if data is pre-fetched, then which data is pre-fetched:

- whether data is shared between clients (and replicated) or owned by one client at a time (and must be transferred)

- how concurrent access and update is controlled and supported

- whether garbage collection or explicit deletion is allowed

- whether references to objects can be taken out of the systems' control

Several of these issues are classically and widely discussed in the database literature in the context of distributed databases. We choose to focus instead on a C or C++ programmer's view, by extending the heap of record-like data objects and pointers to cover a global addressing space potentially larger than the entire multicomputer memory (that is, 2GByte in the case of a 128 cell, 16 MByte per cell machine). The desired semantic model is then similar to that of multiple processes running in a single processor, with a very large shared heap. The multiple processes are however tasks distributed among the multicomputer's cells.

Programmers are willing to have relatively weak data coherence and concurrency control in such a system, but to provide performance we must maximise potential concurrency and minimise the cost of sharing or transferring data around the distributed memory system.

Several of the listed issues affect performance of the shared heap utility. In particular **caching**, *space recovery*, *pre-fetching*, and *data blocking* are significant. To investigate these issues, we have implemented a prototype shared heap service with client-side caching and replicated objects to allow us to vary and investigate the effects of the interacting factors of

- demand fetching only, versus pre-fetching

- methods of prediction for pre-fetching

- storage recovery algorithms in the client cache

The behaviour of the shared heap under these variable factors obviously depends greatly on the pattern of data accesses made by any particular application program. To explore these factors in an idealised experiment we use two tree-traversing programs that are based on real applications.

# 2 The shared heap storage manager

## 2.1 Object identifiers and objects

Every object in the shared heap is referred to by a unique *object identifier* that acts as a global system-wide pointer. The object identifier can be used by any process in possession of it to access the object it identifies. Object identifiers can be assigned to and compared in programs, stored as data in objects or kept as local data in a program.

Objects have a structure imposed upon them, in that they consist an untyped *data part* and a *reference part*. The size of each part is fixed when the object is created. The data part contains byte data for the application program's use as program data. The reference part contains references to other objects in the heap. Using the reference fields as pointers to other objects, the programmer can construct arbitrary graph data structures, just as with ordinary objects in the local memory of a processor.

## 2.2 Managing the shared heap

The shared heap is managed by a *Global Object Manager* (GOM) which manages heap space across one or more *server* processors, whose memory is dedicated to storing the heap. Each server process runs a task which is part of the GOM. Each object is managed by some particular server, and its value is always present as the *master copy* of the object in that server's local memory. (The mapping of object identifier to server is many to one, but fixed). The object may also be replicated in any number of client memories.

Individual *client* processors set aside a portion of their local memory as an *object cache* in which to store replicas of heap objects and supporting data structures. The client memory is managed on demand from the user's application program. The user program makes library calls to perform object creation, read access, and write access (update). The library automatically requests objects to be *fetched* from the Global Object Manager, creates space in its cache if necessary to store the newly fetched objects, and writes back updated objects to the Global Object Manager as necessary.

The interaction between client and object managers is by message passing. The Global Object Manager is in fact only the collection of server tasks; a client initially sends a request to a particular server, and the server may pass requests to other servers to service the request. Requests which refer to a known object identifier are sent directly to the server responsible for that object.

**object creation**

Creating objects in the shared heap is necessarily more expensive than allocating a heap object in local memory. The client interacts with the Global Object Manager to get a unique *object identifier* that refers to newly allocated space in one of the server processors.

Upon receiving the identifier the client then allocates local space for the object and updates the *Local Object Manager* (LOM) *Resident Object Table*. This table (the ROT) maps the object identifier to local storage address.

**object access and update**

When a client attempts to access an object, the Local Object Manager looks up the object identifier in the ROT[1]. If there is a local replica in the cache, the data is read and returned or updated as required, and the table updated if necessary to maintain Least Recently Used information. If there is no cached local replica, the LOM requests it from the GOM using the object identifier by which the access was attempted. The GOM at the server in whose space the object's master copy resides then supplies a replica in reply. The requesting call does not complete until the requested object is received.

A client's Local Object Manager similarly converses with the GOM when its cache becomes full. Space is recovered by discarding objects. Clean replicas can be safely discarded, but those which have been updated must return their value to the GOM to enable updating of the master copy.

## 2.3 Concurrency control

Concurrent access to store objects by multiple clients can be managed in two ways. Most simply a semaphore which is attached to each object can be utilised as a locking mechanism to ensure that at most one client process can access the object at any time. The semaphore is managed by the server for the object and is costly in that each Wait or Signal operation on the semaphore requires an exchange of messages between the local and global managers. This mechanism is however suited to applications that rarely update shared objects, or which can separate the computation into phases of sole ownership and update, followed by phases of read-only sharing. The semaphores need only be used between phases to synchronise the use of objects safely.

Alternatively, if the computation is easily separable into phases in which objects are either non-shared or not updated, then such phases can be separated by explicit barrier synchronisation commands that also write back all updated objects. This second mechanism has been used in our sample applications not only because they are amenable to it, but also because the larger number of client-server synchronisations generated by the first mechanism is likely to be unduly favourable to the pre-fetching mechanism being examined.

## 2.4 Pre-fetching

Many data structures have common patterns of usage, that is, the elements in the structure are often accessed in similar patterns. One of the most common usage patterns is to follow one or more of the pointers contained in a node soon after that node is accessed. For example, a linked list is often traversed in a pattern consisting of examining the value of the current node and then following the pointer to the next node. This pattern occurs as part of most list operations – search, duplicate deletion etc. Similarly, when traversing binary trees it is likely that at least one of the pointers in a node will be followed soon after the node is first examined.

---

[1] For the initial access a search of the ROT is required. An index yielded by this search is then stored in the program's object reference value, allowing subsequent operations to access the ROT entry directly.

We have attempted to generalise this common pattern to the arbitrary directed graphs formed by the objects stored in our shared heap. From this generalisation we sought to improve the performance of the fetch operation by the pre-fetching of objects likely to be accessed next. A pre-fetching method will attempt to fetch the next object from the server while the client is processing a previously fetched object, instead of the client blocking whenever a freshly requested object's reference field is followed.

As described above, the procedure followed when a client attempts to access an object for which it holds no replica is that a request is sent to the GOM, which services it by supplying a copy of the required object. The pre-fetch mechanism implemented within our global storage manager attempts to preempt some future fetch requests. When a request for an object is satisfied in the normal manner the desired object is forwarded to the client. Optionally, the server then queues a request on behalf of the client for those objects referred to in the fields of the requested objected just sent. This occurs transitively to a specified depth, and we refer to this as the *pre-fetch depth*. Thus, if the client process was to subsequently require any of the objects whose references are contained in the reference part of the requested object, these will have at worst already been queued for sending, and at best have already arrived at the client. This mechanism gives the possibility of reducing both the number of fetch requests and the average time required to satisfy those that do occur.

The clients' pattern and frequency of access determines whether each new access incurs small delay, being met by the pre-fetch from a previous real request, or whether the server delays induced by serving unnecessary pre-fetch requests outweigh the benefits to the client.

The client side can suffer from more subtle costs as well. These include quickly filling the local memory necessitating more frequent clearances, and overhead time spent receiving extra objects. At a more general level there is also the problem of generating extra network traffic.

### pre-fetching priority

An interesting issue regarding pre-fetch requests generated by servers is their priority in relation to client generated requests. If a server has both pre-fetch requests and real requests to service which should be attended to first? Two implementations of pre-fetching have been tested, one in which the pre-fetch requests have a priority equal to that of real requests (termed *high priority pre-fetching*), and one in which pre-fetch requests are only serviced in the absence of real requests (called *low priority*).

## 3 Example applications

We have tested the performance of the shared global heap with two applications : insertion and search in binary trees and N-body calculations with oct-trees.

### binary tree applications

The binary tree tests include a tree based sorting algorithm. How to utilise the power of a machines such as the AP1000 when sorting is a common problem that is relevant to many larger applications including document indexing and speech recognition. The application consists of two distinct phases: an insertion phase, and a traversal phase. Several thousand randomly generated keys are inserted into a binary tree during the insertion phase, an in-order traversal of the tree then yields the sorted key sequence. A different usage pattern over the same data structure was utilised as a control measure to determine the generality of the binary tree results. This second pattern was generated by constructing the tree as just mentioned and proceeding to search it for 3000 randomly selected keys some of which were not present.

### N-body oct-tree application

The second application is a modified form of the N-body problem [3, 4]. A physical system consisting of a large number of particles freely interacting with each other is simulated by calculating forces and applying approximate accelerations for small time increments, updating the particles' positions, and repeating this process for each time step. In the initial phase a large number of particles (representing stars, for instance) are generated with a range of masses and positions in 3 dimensions. The main loop repeats force calculations, by summing the force on each particle due to every other particle, followed by updating the position of every particle. The force summation is (naively) of order $N^2$ in the number of particles.

A substantial decrease in computation time can be made by approximating the effect of a group of particles that are relatively far away from a particle of interest. The effect of the group is approximated by the effect of a single particle with the sum of their masses placed at the group's centre of mass. The approximation can be efficiently implemented with the aid of a spatial oct-tree structure, representing combined masses for each group at the internal nodes of the tree. In practice the use of this approximation reduces the computation time to approximately $O(N \log N)$ [3]. A spatial oct-tree has the general form of a node being an approximation for the masses in a volume of space. If this node represents more than a single mass, it has eight children each being the approximation for one octant of their parent's volume.

The programming of the oct-tree represents a challenging task in distributed memory machines if the programmer needs to manage distribution explicitly. An alternative is the creation of a globally linked data structure on a shared heap. All processes are then able to transparently access those portions of the oct-tree required for their calculations and update nodes for which they are responsible.

Our initial experiments traverse the data structures required for the oct-tree approximation method, without in fact applying the approximation. The calculation phase therefore traverses the entire approximation tree for each particle it updates.

For this example, the application has two phases: a structure generation phase, involving tree and list building by one processor, and a calculation phase. During the later phase several processors each assume responsibility for a portion of the particle list and proceed in parallel to traverse the approximation tree in order to calculate the forces acting upon each of the particles in their list.

| pre-fetch depth | create | clear | fetch | execute |
|---|---|---|---|---|
| 0 | 10.5 | 18.7 | 78.3 | 163.2 |
| 1 | 10.6 | 20.2 | 68.1 | 152.9 |
| 2 | 10.5 | 18.3 | 68.0 | 141.5 |

Table 1: breakdown of execution times for binary tree traversal using high priority pre-fetch

## 4 Performance tests & Discussion

We have performed several experiments to determine the value, if any, of pre-fetching, and to expose associated costs. The pre-fetch depth is a selectable parameter of the server in these experiments. The initial experiment involved the binary tree construction and traversal described in §3, and yielded encouraging results. The tree consisted of 6000 nodes with integer keys inserted in random order and then traversed inorder to gain the sorted key sequence. The system consisted of a single server supporting a single client in this application, with high priority pre-fetching. Table 1 gives a breakdown of the client execution times recorded. Note the significant improvement in the total time expended as a result of pre-fetching to a depth of one, and a further improvement with a depth of two. Given these encouraging results the depth of pre-fetching was extended. Figure 1 illustrates the results obtained and indicates a graceful performance degradation for pre-fetching beyond the optimum depth of four.

It was noted in §2.4 that one of the possible disadvantages of pre-fetching was the extra processing load placed upon the servers. The next experiment was run to determine to what extent the extra load would be noticeable in a multi-client environment. A single server was given the task of supporting one to four clients each performing the binary tree construction and traversal of the first experiment. This was done for the control case of no pre-fetching, and for low priority pre-fetching to a depth of two. Figure 2 indicates that the extra server workload can quickly overcome the benefits of pre-fetching, resulting in this case with the addition of the third

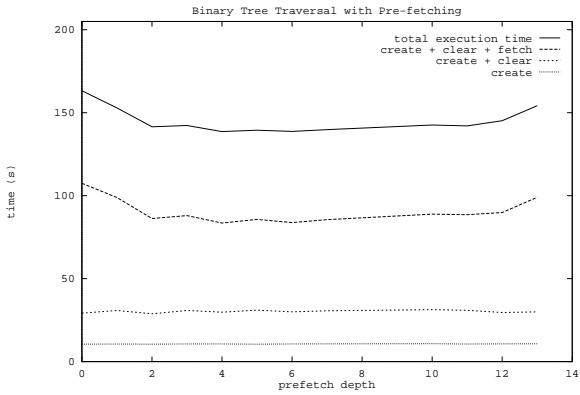

Figure 1: pre-fetch depth in binary tree traversal

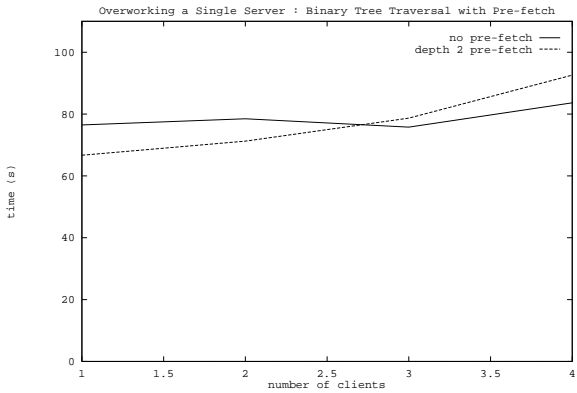

Figure 2: multiple clients — binary tree application

| pre-fetch depth | 0 | 1 | 2 | 3 | 5 |
|---|---|---|---|---|---|
| # untouched | 0 | 0 | 0 | 292 | 1140 |

Table 2: unused pre-fetched objects : binary tree traversal

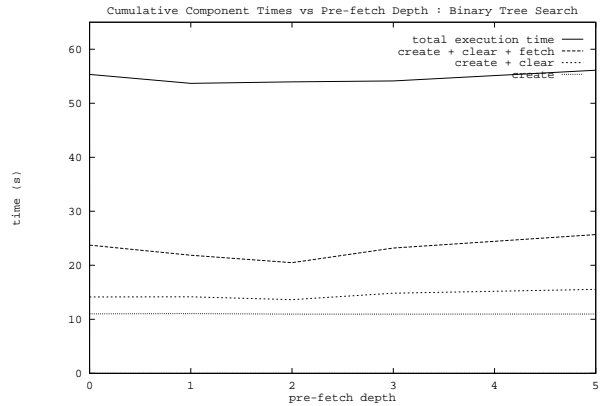

Figure 3: search in the binary tree application

client causing pre-fetch performance to become inferior.

To test the generality of the results gained using the binary tree traversal application, a test similar to the initial experiment was run using the binary tree search application. The tree was constructed as before, but the second phase of computation consisted of searching the tree for three thousand random keys. The second phase usage pattern in this case is quite different from the earlier experiment in that there is no guarantee that any pre-fetched item will ever be accessed, and the top few nodes of the tree will visited frequently as opposed to twice in the tree traversal. Whilst the results (see Figure 3) show some improvement for the first two levels of pre-fetching, it is less significant than that found in Table 1, and disappears relatively quickly. Table 2 indicates that the performance drop off as the level of pre-fetching increases is mirrored by the number of shared heap objects which are discarded from the client's memory untouched (ie: which were pre-fetched in but discarded, due to lack of space, without being accessed).

Two experiments were run using the oct-tree N-body simulation. One thousand stars were inserted into an oct-tree, yielding a tree of approximately 6000 nodes. The tree was then traversed by the client node to calculate forces acting upon each of 100 stars which were then updated with new positions and velocities. The initial oct-tree experiment utilised the same high priority pre-fetch mechanism as the initial binary tree experiment. The changes in time are evidently due to the heap access — the computation times are unchanged. In this case it can be seen that pre-fetching (levels 1, 2 or 3) actually *increases* the overall time compared to fetching only on request (level 0).

The second oct-tree experiment involved the same application but used the server stratgey of low priority pre-fetching. This sought to determine if the pre-fetch requests were forming a bottleneck, slowing down the servicing of real fetch requests. The results in figures 4 and 5 show that the low priority pre-fetching actually had worse overall performance at a depth of one, and whilst it improved marginally at depths two and three, it was still poorer than the high pri-ority pre-fetch in all cases.

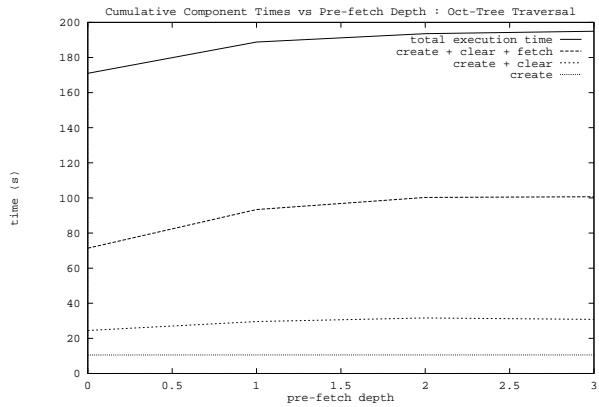

Figure 4: high priority pre-fetching for oct-tree traversal

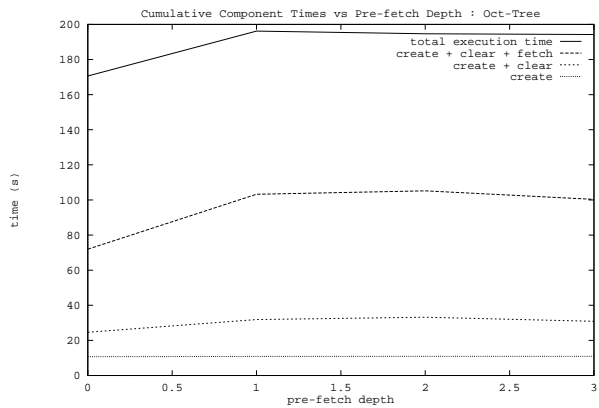

Figure 5: low priority pre-fetching for oct-tree traversal

## 5 Conclusion

Our experience with implementing the utility and programming these applications show that the distributed heap store utility is beneficial to programmers. Subjectively, its performance is acceptable on a low latency message-passing distributed memory system.

The experimental results show that pre-fetching is a beneficial strategy in the management of the global heap. Most benefit (of the order of 10% of data-fetching time) is shown for small pre-fetch levels of 1 or 2 in the specific binary tree applications measured. Some applications (such as the oct-tree traversal) show no gain in performance for pre-fetching. This might be explained because this structure has a very large fan-out at each node (8 children compared to the 2 children in the binary tree), and traversal is depth-first in both cases. Many of the nodes that are pre-fetched will not be used in the near future, and will cause the client cache to fill prematurely, with more frequent expensive space-clearing operations as a result.

The experimental results support the intuitive hypothesis that the performance of pre-fetching depends on the application data structures and on the pattern of access to the structure. However, many applications have different access patterns in different phases of the computation, such as searches in one phase, traversal in another. The best overall performance will be got by using different pre-fetch behaviour for this data structure in each phase. For this reason we believe that the pre-fetching level should be determined differently for classes of objects, and should be specified by the application dynamically. This is in opposition to other workers in the field: this behaviour is often seen as being a static property of the object alone, or a property of the class of objects, as in the Pool Managers of Mneme [2].

We hypothesized that comparing the strategy of giving higher priority to actual requests than pre-fetch requests would perform better than the strategy of giving equal priority to actual requests and pre-fetch requests. The experiment of comparing these strategies, using the oct-tree application, where the large fan-out of the tree creates a large number of pre-fetch requests, did not support this hypothesis. Further experimentation is needed with priority strategies and other more intelligent management of the requests.

## Acknowledgements

The ANU-Fujitsu CAP program provided access to the Fujitsu AP1000 computer at the Department of Computer Science, ANU, for this work.